\documentclass[showpacs,preprintnumbers,amsmath,amssymb]{revtex4}

\usepackage{graphicx}% Include figure files
\usepackage{dcolumn}% Align table columns on decimal point
\usepackage{bm}% bold math
\usepackage{url}
\usepackage{epsfig}
\usepackage{multirow}
\usepackage{tensor}
\usepackage{empheq}
\usepackage{amsmath}
\usepackage{color}
\definecolor{myblue}{rgb}{.8, .8, 1}

\def\be{\begin{equation}}
\def\ee{\end{equation}}
\def\ba{\begin{eqnarray}}
\def\ea{\end{eqnarray}}

\newcommand{\fr}[2]{\frac{#1}{#2}}
\def\kk{\vec{k}}
\def\m{\rm{m}}
\def\D{\rm{D}}
\def\DE{\rm{DE}}
\def\Omo{\Omega_{\rm{m}0}}
\def\Obo{\Omega_{\rm{b}0}}
\def\Omz{\Omega_{\rm{m}}(z)}

\def\cdm{\rm{CDM}}

\def\hMpc{\rm{h\,Mpc^{-1}}}
\def\hiMpc{\rm{h^{-1}\,Mpc}}
\def\eq{\rm{eq}}
\newcommand{\oo}{\omega_{0}}
\newcommand{\oa}{\omega_{a}}

\newcommand{\MNRAS}{Mon.\ Not.\ Roy.\ Astron.\ Soc.}

\def\ga{\mathrel{\raise.3ex\hbox{$>$\kern-.75em\lower1ex\hbox{$\sim$}}}}
\def\la{\mathrel{\raise.3ex\hbox{$<$\kern-.75em\lower1ex\hbox{$\sim$}}}}
\begin{document}

%\begin{titlepage}

\leftline{KIAS-P14054}

\title{Breaking CMB degeneracy in dark energy through LSS}

%\\

\author{Seokcheon Lee}

\affiliation{School of Physics, Korea Institute for Advanced Study, Heogiro 85, Seoul 130-722, Korea}

%\date{May 11, 2010}% It is always \today, today,

\begin{abstract}
The cosmic microwave background (CMB) and large scale structure (LSS) are complementary probes to investigate the early and late time universe. After the current accomplishment of the high accuracies of CMB measurements, accompanying precision cosmology from LSS data is emphasized. We investigate the dynamical dark energy (DE) models which can produce the same CMB angular power spectra as that of the $\Lambda$CDM model with less than a sub-percent level accuracy. If one adopts the dynamical DE models using the so-called Chevallier-Polarski-Linder (CPL) parametrization, $\omega \equiv \omega_{0} + \omega_{a}(1-a)$, then one obtains models $(\omega_{0},\omega_{a}) = (-0.8,-0.767), (-0.9,-0.375), (-1.1,0.355), (-1.2,0.688)$ named as M8, M9, M11, and M12, respectively. The differences of the growth rate, $f$ which is related to the redshift space distortions (RSD) between different DE models and the $\Lambda$CDM model are about 0.2 \% only at z=0. The difference of $f$ between M8 (M9, M11, M12) and the $\Lambda$CDM model becomes maximum at $z \simeq 0.25$ with -2.4 (-1.2, 1.2, 2.5) \%. This is a scale-independent quantity. One can investigate the one-loop correction of the matter power spectrum of each model using the standard perturbation theory in order to probe the scale-dependent quantity in the quasi-linear regime ({\it i.e.} $k \leq 0.4 \hiMpc$). The differences in the matter power spectra including the one-loop correction between M8 (M9, M11, M12) and the $\Lambda$CDM model for $k= 0.4 \hiMpc$ scale are 1.8 (0.9, 1.2, 3.0) \% at $z=0$, 3.0 (1.6, 1.9, 4.2) \% at $z=0.5$, and 3.2 (1.7, 2.0, 4.5) \% at $z=1.0$. The bigger departure from $-1$ of $\omega_{0}$, the larger the difference in the power spectrum. Thus, one should use both the RSD and the quasi-linear observable in order to discriminate a viable DE model among a slew of models which are degenerated in CMB. Also we obtain the lower limit on $\oo > -1.5$ from the CMB acoustic peaks and this will provide the useful limitation on phantom models.

\end{abstract}


%\end{titlepage}

\maketitle

%%%%%%%%%%%%%%%%%%%%%%%%%%%%%%%%%%%%%%%%%%%%%%%%%%%%%%%%
\section{Introduction}
\setcounter{equation}{0}
%%%%%%%%%%%%%%%%%%%%%%%%%%%%%%%%%%%%%%%%%%%%%%%%%%%%%%%%

Both the cosmic microwave background (CMB) and the large scale structure (LSS) in the Universe have been used to constrain cosmological parameters. Especially, the growth history of the matter fluctuation from LSS is used to reveal the properties of dark energy (DE). Although CMB anisotropies furnish the limited information about the DE on their own, CMB constraints on the geometry and the matter (radiation) content of the Universe play a crucial role in probing DE when combined with low redshift surveys. CMB data supply measurements of the observed angular size of the sound horizon at recombination $\theta_{s} = r_{s} / d_{A}^{(c)}$ from the angular location of the acoustic peaks to better than 0.1 \% precision at 1 $\sigma$ \cite{Planck}. Even though the sound horizon at the time of last scattering, $r_{s}(z_{\ast})$, is insensitive to the properties of DE, the comoving angular diameter distance at which we are observing the fluctuations, $d_{A}^{(c)}(z_{\ast})$, does depend on the properties of DE.

CMB also provides the best way of fixing the amplitude of cosmological fluctuations on the largest scales \cite{9601170, 9607060}. In addition to this, as a secondary anisotropies the different amounts of potentials decay caused by different DE models lead to the net energy change of photons called the Integrated Sachs-Wolfe (ISW) effect. The alternative normalization is $\sigma_8$, the {\it rms} linear matter fluctuation in spheres of radius 8 $\hiMpc$ inferred from abundances of clusters. However, this scale is not sufficiently large to remove the non-linear effect and fluctuations at these scales are still well inside the horizon to depend on its evolution.

Redshift space distortions (RSD) are the consequence of peculiar motions on the measurement of the power spectrum from a galaxy redshift survey. On large scales, coherent bulk flows bound to a over density out of voids are coherent towards the central mass which lead to an enhancement in the density inferred in the redshift space. The enhancement of the power spectrum due to RSD, under the linear perturbation theory assumption with the plane parallel approximation is given by $P_{s}(k,\mu) = (1+ \beta \mu^2)^2 P_{r}(k,\mu)$. $\beta$ is a so-called the RSD parameter defined as $\beta = f / b$ where $f = d \ln \delta / d \ln a$ is the growth rate and $b$ is the bias factor \cite{Kaiser}.

Our main interest is to understand the dark-energy effect on the matter power spectrum in a quasi-linear regime. Analytical solutions are presented for the dynamical DE model parameterized so-called Chevallier-Polarski-Linder (CPL) parametrization, $\omega \equiv \oo + \oa(1-a)$ \cite{CPL}. Each dynamical model can produce the same CMB power spectrum as those of $\Lambda$CDM model by obtaining the proper $\oo$ and $\oa$ fixing all other cosmological parameters. Similar approach for the numerical simulation has been investigated \cite{07040312} and the approximate approach using the standard perturbation theory (SPT) for the one-loop correction matter power spectrum has also been studied \cite{08061437}.

In the next section, we obtain the proper values of ($\oo$, $\oa$) to produce the same angular size of the sound horizon power as that of $\Lambda$CDM model. We compare the CMB power spectra of models. In section 3, we compare the predicted values of RSD for corresponding models. In section 4, we obtain the one-loop matter power spectrum of each model using SPT and compare it with that of $\Lambda$CDM model. We conclude in the last section. In the appendix, we extend the models including ones will be possibly ruled out in future survey.

%%%%%%%%%%%%%%%%%%%%%%%%%%%%%%%%%%%%%%%%%%%%%%%%%%%%%%%%
\section{Dark Energy and Cosmic Microwave Background}
\renewcommand{\theequation}{2-\arabic{equation}}
\setcounter{equation}{0}
%%%%%%%%%%%%%%%%%%%%%%%%%%%%%%%%%%%%%%%%%%%%%%%%%%%%%%%%

The CMB is a power window to probe the early universe. At the last scattering surface, $z_{\ast}$ where the photons interact with matter for the last time, it shows tiny temperature fluctuations that correspond to slightly different densities, representing the seeds of LSS. The pressure of the photons tends to erase temperature anisotropies, whereas the gravitational attraction of the baryons makes them tend to collapse. These two effects compete to create acoustic oscillations with CMB peak structure. One calls the characteristic angular size of the fluctuation in the CMB as the acoustic scale. It is determined by the sound horizon at the last scattering, $r_{s}(z_{\ast})$ and the comoving angular diameter distance, $d_{A}^{(c)}(z_{\ast})$. We adopt CPL parametrization of DE equation of state, $\omega$. The acoustic angular size is defined by
\be \theta_{s}(z_{\ast}) = \fr{r_{s}(z_{\ast})}{d_{A}^{(c)}(z_{\ast})} \, , \label{thetas} \ee
where
\ba r_{s}(z_{\ast}) &=& \int_{0}^{t_{\ast}} \fr{ c_{s} dt}{a} = \fr{c}{\sqrt{3} H_0} \int_{z_{\ast}}^{\infty} \fr{dz'}{\sqrt{1+R(z')} E(z')} \, , {\rm where} \,\, R \equiv \fr{3 \rho_{b}}{4 \rho_{\gamma}} \, , \label{rs} \\
d_{A}^{(c)}(z_{\ast}) &\equiv& (1+z) d_{A}(z_{\ast}) = \fr{c}{H_0} \int_{0}^{z_{\ast}} \fr{dz'}{E(z')} \, , \label{dAc} \\
E(z) &\equiv& \fr{H}{H_0} \label{Ez} \\
&=& \sqrt{\Omo (1+z)^3 + \fr{\Omo}{1+z_{\eq}} (1+z)^4 + (1-\Omo\fr{2+z_{\eq}}{1+z_{\eq}}) (1+z)^{3(1+w_0+w_a)} \exp[-3w_a(\fr{z}{1+z})]} \nonumber \, . \ea
where $H_0$ is the present value of the Hubble parameter, $\Omo$ is the present value of the matter energy density contrast, and $z_{\eq}$ is the
matter and the radiation equality epoch. $\theta_{s}(z_{\ast})$ is measured by the positions of the peaks but not by their amplitudes and this it is quite robust. Also, $\theta_{s}(z_{\ast})$ is tightly constrained from the observation, it can be safely used to constrain the cosmological parameters. From Eqs. (\ref{thetas})-(\ref{Ez}), one can find that $d_{A}^{(c)}(z_{\ast})$ depends on $\omega$ and so does $\theta_{s}(z_{\ast})$. If one keeps all other cosmological parameters fixed except $\omega$, then one is able to obtain the viable values of ($\oo, \oa$) which can produce the same $\theta_{s}(z_{\ast})$ as that of $\Lambda$CDM model. We find the viable models and show their ($\oo, \oa$) values in Table. \ref{tab1}.

%%%%%%%%%%%%%%%%%%%%%%%%%%%%%%%%%%%%%%%%%%%%%%
\begin{center}
\begin{table}
\begin{tabular}{ |c||c|c|c|  }
 \hline
 Models & ($\oo, \oa$) & $\sigma_8$ & $t_{0}$ (Gyr) \\
 \hline
 M8   & (-0.8,-0.767)    & 0.850 & 13.40\\
 M9   & (-0.9,-0.375)  & 0.848   & 13.43\\
 $\Lambda$CDM & (-1.0,0.0) & 0.845 & 13.46 \\
 M11  & (-1.1,0.355) & 0.842 &  13.50 \\
 M12  & (-1.2,0.688)  & 0.837 & 13.54 \\
 \hline
\end{tabular}
\caption{$\sigma_8$ and the age of the Universe in (Gyr), $t_0$ for CMB degenerated DE models.}
\label{tab1}
\end{table}
\end{center}
%%%%%%%%%%%%%%%%%%%%%%%%%%%%%%%%%%%%%%%%%%%%%%%%
We put $\Omo = 0.3$, $\Obo = 0.0462$, $\Omega_{\gamma 0} = 5.04 \times 10^{-5}$, $H_0 = 70$h km/sec/Mpc, $z_{\eq} = 3513$, and $z_{\ast} = 1089.73$. This set of parameters produces $\theta_{s}(z_{\ast}) = 0.0105$ for the $\Lambda$CDM. If we vary the value of $\oo$ from -1.2 to -0.8 to obtain the same value of $\theta_{s}(z_{\ast})$ as $\Lambda$CDM, then we obtain $\oa$ values as given in the Table. \ref{tab1}. We label each model ($\oo,\oa$) = (-0.8,-0.767), (-0.9,-0.375), (-1.1,0.355), and (-1.2,0.688) as M8, M9, M11, and M12, respectively.

\begin{figure}
\centering
\vspace{1.5cm}
\begin{tabular}{cc}
\epsfig{file=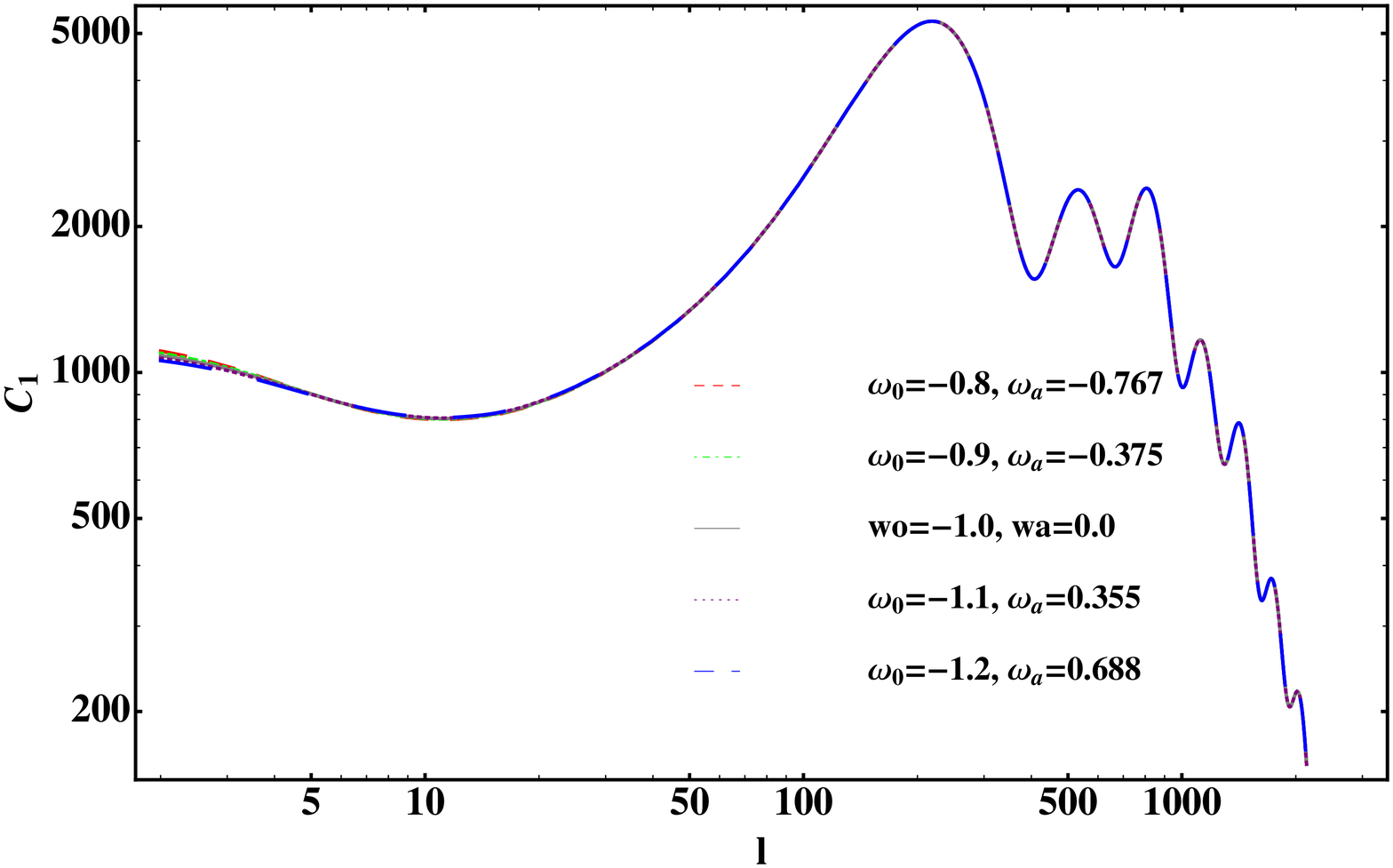,width=0.45\linewidth,clip=} &
\epsfig{file=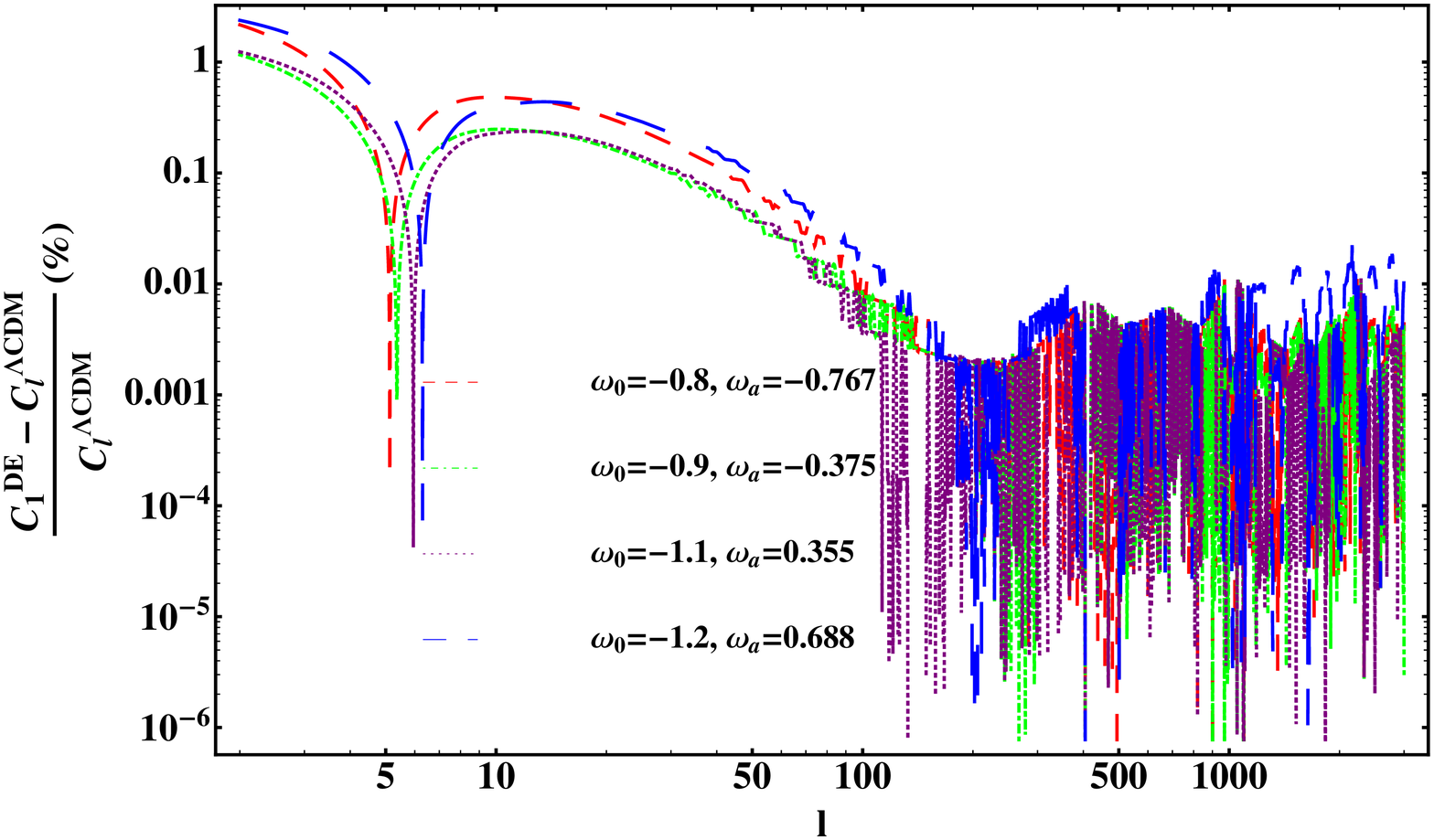,width=0.45\linewidth,clip=} \\
\end{tabular}
\vspace{-0.5cm}
\caption{CMB angular power spectra and their differences. left) CMB angular power spectra for M8 (dashed), M9 (dot-dashed), $\Lambda$CDM (solid), M11 (dotted), and M12 (long-dashed), respectively. right) Differences between CMB angular power spectrum of $\Lambda$CDM model and that of M8 (dashed), M9 (dot-dashed), M11 (dotted), and M12 (long-dashed), respectively.} \label{fig1}
\end{figure}

We show the CMB angular power spectrum of each model and its difference from $\Lambda$CDM model one in Fig. \ref{fig1}. As we expect, the CMB angular power spectra between models are almost same to one another as shown in the left panel of the Fig. \ref{fig1}. Dashed, dotdashed, solid, dotted, and long-dashed lines correspond ($\oo,\oa$) = (-0.8,-0.767), (-0.9,-0.375), (-1.0, 0), (-1.1,0.355), and (-1.2,0.688), respectively. If we investigate the differences of CMB power spectra between DE models and $\Lambda$CDM model, then they are less than 1 \% for all model when $l \geq 5$. The difference in the large scale is due to the integrated Sachs Wolfe (ISW) effect caused by gravitational redshift occurring between the surface of last scattering and the present epoch. The different DE models produce the different large-scale gravitational potential energy wells and hills evolution and they cause changing the energy of photons passing through them. This is shown in the right panel of Fig. \ref{fig1}. Dashed, dotdashed, dotted, and long dashed lines correspond the angular power spectra differences between $\Lambda$CDM and ($\oo,\oa$) = (-0.8,-0.767), (-0.9,-0.375), (-1.1,0.355), and (-1.2,0.688), respectively. Observationally it is impossible to distinguish the CMB angular power spectra for these different models at large scale due to the cosmic variance.

CMB with the different DE models also provides the different normalization at large scale as \cite{9906174, 10102291}
\be P(k,a) = A k^{n_s} T(k)^2 \Bigl(\fr{D(a)}{D_{0}} \Bigr)^2 \equiv 2 \pi^2 \delta_{H}^2 \Bigl(\fr{c}{H_0}\Bigr)^{n_s+3} k^{n_s} T(k)^2 \Bigl(\fr{D(a)}{D_{0}} \Bigr)^2 \label{Pka} \, , \ee
where $A$ is the normalization, $n_s$ is the spectral index of the primordial adiabatic density perturbations, $T$ is the transfer function, $D(a) (D_0)$ is the linear growth factor at $a (a=1)$, and $\delta_{H}$ is the horizon crossing amplitude. From CMB observation, one can extract $\delta_{H}$ for the different DE models. In other word, $\delta_{H}$ can be a function of $\omega$. However, theoretically this value is determined from the specific inflation model and thus we use the same value of $\delta_{H}$ ({\it i.e.} $A$) for the matter power spectrum analysis in Sec. 4. This also explains why we obtain the slightly different values of $\sigma_8$ for the different models.
\be \sigma_{b}^2(a) \equiv \Bigl\langle \Bigl|\fr{\delta M}{M(R,a)}\Bigr|^2 \Bigr\rangle = \fr{1}{2\pi^2}\int_{0}^{\infty} k^2 P(k,a) \Bigl|W(kR)\Bigr|_{R=8\hiMpc}^2 dk \label{sigma8} \, . \ee
Even though we use the same values of $\delta_{H}$ and $n_s$ for the different DE models, one obtains the different $T(k)$s and these cause the slight different values of $\sigma_8$ as shown is table \ref{tab1}.

%%%%%%%%%%%%%%%%%%%%%%%%%%%%%%%%%%%%%%%%%%%%%%
%%%%%%%%%%%%%%

%%%%%%%%%%%%%%%%%%%%%%%%%%%%%%%%%%%%%%%%%%%%%%%%%%%%%%%%
\section{Dark Energy and Redshift Space Distortions}
\renewcommand{\theequation}{3-\arabic{equation}}
\setcounter{equation}{0}
%%%%%%%%%%%%%%%%%%%%%%%%%%%%%%%%%%%%%%%%%%%%%%%%%%%%%%%%

Although Hubble's law determines the redshift corresponds to true distance, peculiar velocities not associated with the Hubble flow cause distortions in redshift space. These peculiar motions produce two different types of distortion to the matter power spectrum. On small scales, random velocity dispersions in galaxy clusters cause structure to appear elongated with long thin filaments in redshift space point directly back at observer. This is called as ``the finger of God'' effect. On large scales, peculiar velocities of galaxies bound to a central mass during their infall. Peculiar velocities are coherent towards the central mass and cause the deviation of measured redshifts from a pure Hubble's law. This leads to an enhancement in the density inferred in redshift space and called as Redshift space distortions (RSD). The enhancement of the power spectrum due to RSD, under the linear perturbation theory assumption with the plane parallel approximation is given by $P_{s}(k,\mu) = (1+ \beta \mu^2)^2 P_{r}(k,\mu)$ where $P_{s}$ is the matter power spectrum in the redshift-space, $P_{r}$ is one in the real space, $\mu = \hat{k} \cdot \hat{r}$ with $\hat{r}$ denoting the unit vector along the line of sight, and $\beta$ is so-called the RSD parameter defined as $\beta = f / b$ where $f = d \ln \delta / d \ln a$ is the growth rate and $b$ is the bias factor. Since one cannot directly measure the matter power spectrum, one has to investigate the RSD to the matter power spectrum of the galaxies as matter tracer in the galaxy redshift survey. Alternatively, one can use bias free RSD measurement using $f \sigma_8$.

\begin{figure}
\centering
\vspace{1.5cm}
\begin{tabular}{cc}
\epsfig{file=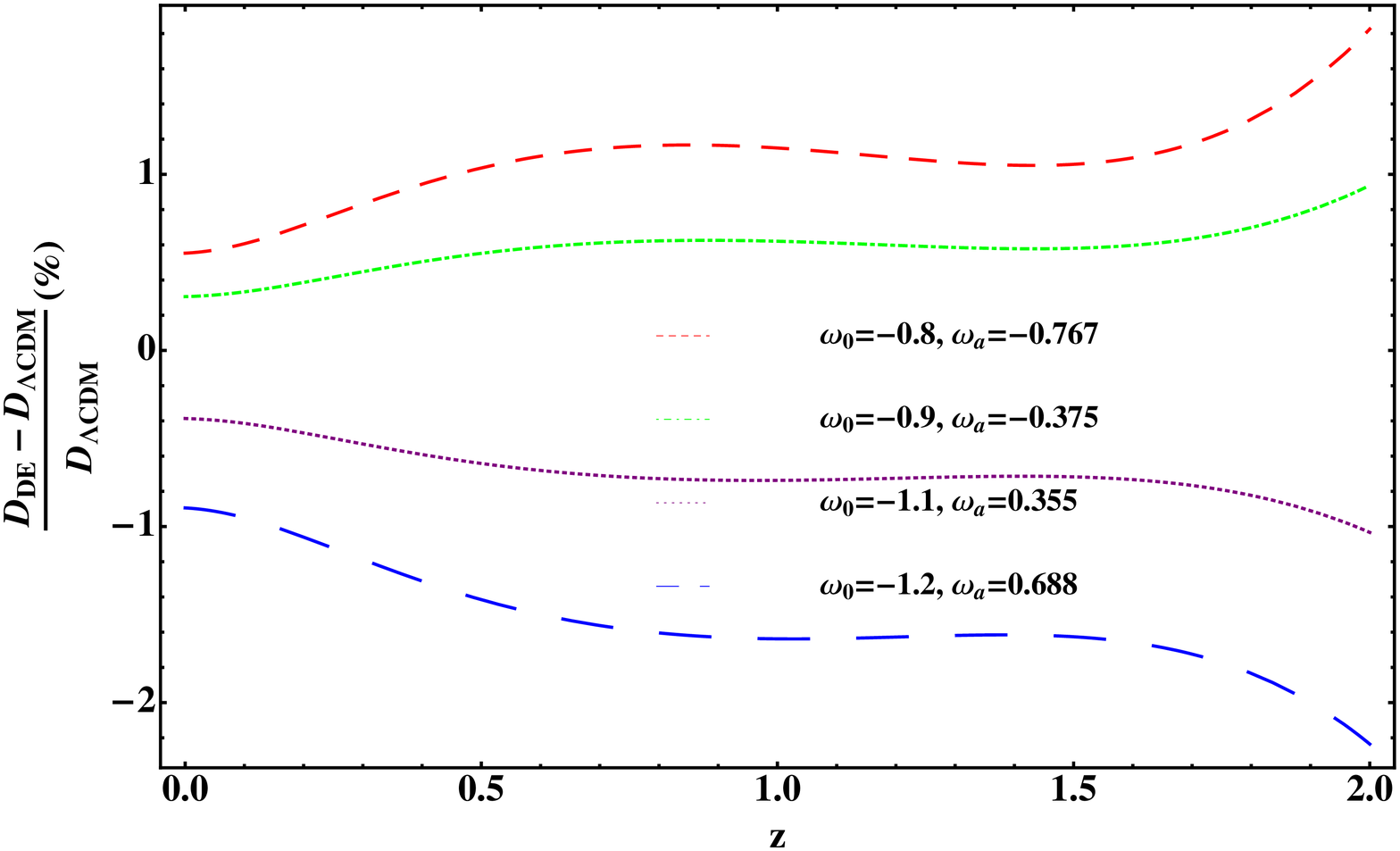,width=0.45\linewidth,clip=} &
\epsfig{file=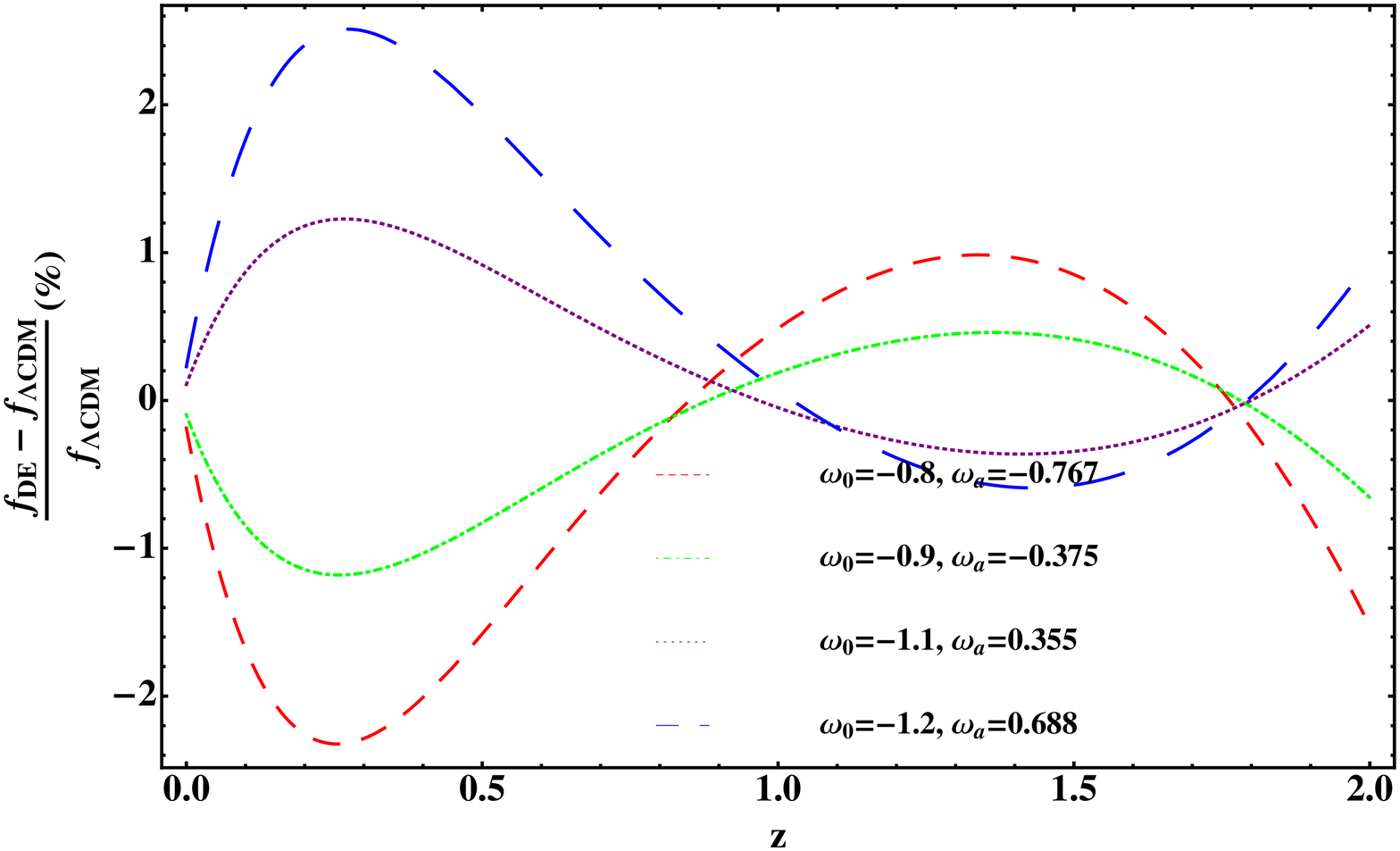,width=0.45\linewidth,clip=} \\
\end{tabular}
\vspace{-0.5cm}
\caption{Differences in the growth factor $D$ and the growth rate $f$ as a function of $z$ left) Differences of growth factors $D$s between DE models and the $\Lambda$CDM. The notation is M8 (dashed), M9 (dot-dashed), M11 (dotted), and M12 (long-dashed), respectively.  right) Differences of growth rates $f$s between DE models and the $\Lambda$CDM. We use the same notation as a).} \label{fig2}
\end{figure}

In sub-horizon scales, one can define both the scale independent matter fluctuation $\delta(k,a) = D(a) \delta(k)$ and its growth rate $f = d \ln D/ d \ln a$ where $D$ is obtained from the linear perturbation theory
\be \fr{d^2 D}{da^2} + \fr{3}{2a} \Bigl(1 - w \Omega_{de} \Bigr) \fr{d D}{da} - \fr{3}{2a^2} \Omega_{m} D = 0 \, , \label{D1eq} \ee where $\Omega_{m} = 1 - \Omega_{de} = \Bigl(1 + (\Omo^{-1}-1) (1+z)^{3(w_0+w_a)} \exp[-3w_a(\fr{z}{1+z})] \Bigr)^{-1}$. $D$ is the sub-horizon scale growth factor. Due to the $\omega$ dependence on $\Omz$, both $D$ and $f$ also depends on $\omega$. However, the differences of $D$ and $f$ between different models are expected to be very small due to the similar background evolutions between them. We show this in Fig. \ref{fig2}.

If we compare the evolutions of the matter energy density contrast, $\Omz$ for different models, then M8 has the biggest $\Omz$ during the cosmic evolution. On the other hand, M12 maintains the smallest $\Omz$ among models. $\Omz$ provides the source term in Eq. (\ref{D1eq}). Thus, one can expect the biggest $D$ for M8 and the smallest one for M12. This is shown in the left panel of \ref{fig2}. The difference of $D$ between M8 (M9, M11, M12) and $\Lambda$CDM is depicted as dashed (dot-dashed, dotted, long-dashed) lines. At the present epoch, $z=0$, the difference of $D$ between M8 (M9, M11, M12) and $\Lambda$CDM is 0.4 (0.25, -0.2, -0.9) \%. Thus, it is impossible to distinguish the different DE models with RSD when the measurement accuracy is bigger sub percent level.  The difference of $D$ becomes maximum at $z \simeq 0.8$ and it is about 1.2 (0.6, -0.7, -1.8) \% for M8 (M9, M11, M12). We also show the differences of $f$ between models in the right panel of Fig. \ref{fig2}. At present epoch, the differences between all models are less 0.2 \%. The difference of $f$ between M8 (M9, M11, M12) and $\Lambda$CDM becomes maximum around $z \simeq 0.25$ with -2.4 (-1.2, 1.2,  2.5) \% deviation. Thus, measurements of $f$ at specific epochs are quite important to probe the DE from RSD.

%%%%%%%%%%%%%%%%%%%%%%%%%%%%%%%%%%%%%%%%%%%%%%
%%%%%%%%%%%%%%

%%%%%%%%%%%%%%%%%%%%%%%%%%%%%%%%%%%%%%%%%%%%%%%%%%%%%%%%
\section{Dark Energy and one-loop Matter Power Spectrum}
\renewcommand{\theequation}{4-\arabic{equation}}
\setcounter{equation}{0}
%%%%%%%%%%%%%%%%%%%%%%%%%%%%%%%%%%%%%%%%%%%%%%%%%%%%%%%%

The standard perturbation theory (SPT) has been widely used to investigate the correction to the linear power spectrum  in a quasi-nonlinear regime. The exact solutions for the Fourier components of the matter density fluctuation $\hat{\delta}(\tau,\kk)$ and the divergence of the peculiar velocity $\hat{\theta}(\tau,\kk)$ has been obtained for general DE models up to third order \cite{LPB}. One can investigate the DE effects on the matter power spectrum including the one-loop correction with these exact solutions.

The equations of motion of $\hat{\delta}(\tau,\kk)$ and $\hat{\theta}(\tau,\kk)$ in the Fourier space are given by
\ba \fr{\partial \hat{\delta}}{\partial \tau} + \hat{\theta} &=& - \int d^3 k_{1} \int d^3 k_2 \delta_{\D}(\vec{k}_{12} - \vec{k}) \alpha(\vec{k}_1, \vec{k}_2) \hat{\theta} (\tau,\vec{k}_1) \hat{\delta} (\tau,\vec{k}_2) \label{massFT} \, , \\
\fr{\partial \hat{\theta}}{\partial \tau} + {\cal H} \hat{\theta} + \fr{3}{2} \Omega_{\m} {\cal H}^2 \hat{\delta} &=& - \fr{1}{2} \int d^3 k_{1} \int d^3 k_2 \delta_{\D}(\vec{k}_{12} - \vec{k}) \beta(\vec{k}_1, \vec{k}_2) \hat{\theta} (\tau,\vec{k}_1) \hat{\theta} (\tau,\vec{k}_2) \label{EulerFT} \, ,
\ea
where $\tau$ is the conformal time, $\vec{k}_{12} \equiv \vec{k}_1 + \vec{k}_2$, $\delta_{\D}$ is the Dirac delta function, ${\cal H} \equiv \fr{1}{a} \fr{\partial a}{\partial \tau}$, $\Omega_{\m}$ is the matter energy density contrast, $\alpha(\vec{k}_1, \vec{k}_2) \equiv \fr{\vec{k}_{12} \cdot \vec{k}_1}{k_1^2}$, and $\beta(\vec{k}_1, \vec{k}_2) \equiv \fr{k_{12}^2 (\vec{k}_1 \cdot \vec{k}_2)}{k_1^2 k_2^2}$.

Due to the mode coupling of the nonlinear terms shown in the right hand side of Eqs. (\ref{massFT}) - (\ref{EulerFT}), one needs to make a perturbative expansion in $\hat{\delta}$ and $\hat{\theta}$ \cite{0112551}. One can introduce the proper perturbative series of solutions for the fastest growing mode $D_{n}$
\ba \hat{\delta}(a,\vec{k}) &\equiv& \sum_{n=1}^{\infty} \hat{\delta}^{(n)} (a,\vec{k})  \label{hatdeltaS} \, , \\
\hat{\theta}(a,\vec{k}) &\equiv& \sum_{n=1}^{\infty} \hat{\theta}^{(n)} (a,\vec{k}) \label{hatthetaS} \, , \ea
where one can define the each order solution as
\ba \hat{\delta}^{(1)} (a,\vec{k}) &\equiv& D_{1}(a) \delta_{1}(\vec{k}) \, , \label{delta1} \\
\hat{\theta}^{(1)} (a,\vec{k}) &\equiv& D_{\theta 1}(a) \theta_{1}(\vec{k}) \equiv -a {\cal H} \fr{d D_1}{da} \delta_{1}(\vec{k})  \, , \label{theta1} \\
\hat{\delta}^{(2)}(a,\vec{k}) &\equiv& D_{21}(a) K_{21}(\vec{k}) + D_{22}(a) K_{22}(\vec{k}) \equiv D_{1}^2 \Biggl[ c_{21}(a) K_{21}(\vec{k}) + c_{22}(a) K_{22}(\vec{k}) \Biggr] \equiv D_{1}^2(a) \delta_{2}(a,\vec{k}) \nonumber \\
&\equiv& D_{1}^2 \int d^3 k_{1} \int d^3 k_2 \delta_{\D}(\vec{k}_{12} - \vec{k}) F_{2}^{(s)}(a, \vec{k}_1,\vec{k}_2) \delta_1(\vec{k}_1) \delta_1(\vec{k}_2) \, , \label{delta22} \\
\hat{\theta}^{(2)}(a,\vec{k}) &\equiv& D_{\theta 21}(a) K_{21}(\vec{k}) + D_{\theta 22}(a) K_{22}(\vec{k}) \equiv D_1 \fr{\partial D_{1}}{\partial \tau} \Biggl[ c_{\theta 21}(a) K_{21}(\vec{k}) + c_{\theta 22}(a) K_{22}(\vec{k}) \Biggr] \nonumber \\ &\equiv&  D_1 \fr{\partial D_{1}}{\partial \tau} \theta_{2}(a,\vec{k})
\equiv -D_1 \fr{\partial D_{1}}{\partial \tau} \int d^3 k_{1} \int d^3 k_2 \delta_{\D}(\vec{k}_{12} - \vec{k}) G_{2}^{(s)}(a,\vec{k}_1,\vec{k}_2) \delta_1(\vec{k}_1) \delta_1(\vec{k}_2) \, , \label{theta22} \\
\hat{\delta}^{(3)}(a,\vec{k}) &\equiv& D_{31}(a) K_{31}(\vec{k}) + \cdots +D_{36}(a) K_{36}(\vec{k}) \equiv D_1^3(a) \Biggl[ c_{31}(a) K_{31}(\vec{k}) + \cdots + c_{36}(a) K_{36}(\vec{k}) \Biggr] \nonumber \\ &\equiv& D_1^3(a) \int d^3 k_1 d^3 k_2 d^3 k_3 \delta_{\D}(\vec{k}_{123} - \vec{k}) F_{3}^{(s)}(a,\vec{k}_1,\vec{k}_2,\vec{k}_3) \delta_{1}(\vec{k}_1) \delta_{1}(\vec{k}_2) \delta_{1}(\vec{k}_3) \, , \label{delta32}\ea
where
\ba c_{2i} &=& \fr{D_{2i}}{D_1^2}\, , \,\,\,\, c_{\theta 2i} = \fr{D_{\theta 2i}}{D_1} \Bigl( \fr{\partial D_{1}}{\partial \tau} \Bigr)^{-1}\, , \,\,\,\, c_{3i} = \fr{D_{3i}}{D_1^3} \, , \label{ci} \\
K_{21}(\vec{k}) &=& -\int d^3 k_{1} \int d^3 k_2 \delta_{\D}(\vec{k}_{12} - \vec{k}) \alpha(\vec{k}_1, \vec{k}_2) \theta_1 (\vec{k}_1) \delta_1 (\vec{k}_2) \, , \label{K21} \\
K_{22}(\vec{k}) &=&  - \int d^3 k_{1} \int d^3 k_2 \delta_{\D}(\vec{k}_{12} - \vec{k}) \beta(\vec{k}_1, \vec{k}_2) \theta_1 (\vec{k}_1) \theta_1 (\vec{k}_2) \, , \label{K22} \\
F_{2}^{(s)}(a,\vec{k}_1,\vec{k}_2) &=& \fr{1}{2} \Biggl[ c_{21} \Bigl( \fr{\vec{k}_{12} \cdot \vec{k}_1}{k_1^2} + \fr{\vec{k}_{12} \cdot \vec{k}_2}{k_2^2} \Bigr) - 2 c_{22} \fr{k_{12}^2 (\vec{k}_1 \cdot \vec{k}_2)}{k_1^2 k_2^2} \Biggr] \label{F2s} \\
&=& c_{21} -2 c_{22} \Biggl(\fr{\vec{k}_1 \cdot \vec{k}_2}{k_1 k_2} \Biggr)^2 + \fr{1}{2} \Bigl(c_{21} -2 c_{22} \Bigr) \vec{k}_1 \cdot \vec{k}_2 \Biggl(\fr{1}{k_1^2} + \fr{1}{k_2^2} \Biggr) \, , \nonumber \\
G_{2}^{(s)}(a,\vec{k}_1,\vec{k}_2) &=& \fr{1}{2} \Biggl[ -c_{\theta 21} \Bigl( \fr{\vec{k}_{12} \cdot \vec{k}_1}{k_1^2} + \fr{\vec{k}_{12} \cdot \vec{k}_2}{k_2^2} \Bigr) + 2 c_{\theta 22} \fr{k_{12}^2 (\vec{k}_1 \cdot \vec{k}_2)}{k_1^2 k_2^2} \Biggr] \label{G2s} \\
&=& - c_{\theta 21} +2 c_{\theta 22} \Biggl(\fr{\vec{k}_1 \cdot \vec{k}_2}{k_1 k_2} \Biggr)^2 - \fr{1}{2} \Bigl(c_{\theta 21} -2 c_{\theta 22} \Bigr) \vec{k}_1 \cdot \vec{k}_2 \Biggl(\fr{1}{k_1^2} + \fr{1}{k_2^2} \Biggr) \, , \nonumber \\
F_{3}^{(s)}(a,\vec{k}_1,\vec{k}_2,\vec{k}_3) &=& \sum_{i=1}^{6} F_{3i}^{(s)}(a,\vec{k}_1,\vec{k}_2,\vec{k}_3) \, , \label{F3s}
\ea
where explicit forms of $F_{3i}^{(s)}$ are given in the appendix of the reference \cite{LPB}. Now we replace $D_{1}(a)$ as $D(a)$

Both the linear and the one-loop power spectra are defined as
\ba P_{1}(a,k) &=& \Biggl( \fr{D(a)}{D_0} \Biggr)^2 P_{11}(k) \, , \label{P1} \\
P_{2}(a,k) &=& \Biggl( \fr{D(a)}{D_0} \Biggr)^4 \Biggl[ P_{22}(a,k) + 2 P_{13}(a,k) \Biggr] \, , \label{P2} \ea where $D_0 = D(a=1)$, $P_{22}$ and $P_{13}$ are obtained as
\ba P_{22}(a,k) &=& 2 \int d^3 q P_{11}(q) P_{11} (|\vec{k}-\vec{q}\,|) \Bigl[ F_{2}^{(s)}(a,\vec{q}, \vec{k}-\vec{q}\,) \Bigr]^2 =
 \fr{(2 \pi)^{-2} k^3}{2} \int_{0}^{\infty} dr P_{11}(kr) \nonumber \\ &\times& \int_{-1}^{1} dx P_{11} \Bigl( k\sqrt{1+r^2-2rx} \Bigr) \Biggl[ \fr{(c_{21}+2c_{22}) r + (c_{21}-2c_{22}) x -2 c_{21} r x^2}{(1 + r^2 - 2rx)} \Biggr]^2 \, , \label{P22k} \\
2 P_{13}(a,k) &=& 6 P_{11}(k) \int d^3 q P_{11}(q) F_{3}^{(s)} (a, \vec{q}, \, -\vec{q}, \, \vec{k}\,) \label{P13k} \\
&=& (2\pi)^{-2} k^3 P_{11}(k) \int_{0}^{\infty} dr P_{11}(kr) \Biggl[ 2c_{35} r^{-2} -\fr{1}{3} \Bigl( 4 c_{31} -8 c_{32} +3c_{33} +24c_{35} - 16 c_{36} \Bigr) \nonumber \\ &-& \fr{1}{3}\Bigl(4 c_{31} -8c_{32} +12c_{33}-8c_{34}+6c_{35} \Bigr)r^2 + c_{33} r^4 + \Bigl(\fr{r^2-1}{r} \Bigr)^3 \ln \Bigl|\fr{1+r}{1-r} \Bigr| \Bigl(c_{35} - \fr{1}{2}c_{33}r^2 \Bigr) \Biggr] \nonumber  \, ,\ea
where $r = \fr{q}{k}$ and $x = \fr{\vec{q} \cdot \vec{k}}{q k}$. Terms with $c_{2i}$ and $c_{3i}$ represent the dark energy effect on the one-loop power spectrum.

We obtain the one-loop power spectra for different DE models by running the camb to obtain the linear power spectrum \cite{camb} using $n_{s} = 0.96$ and $A = 2.1 \times 10^{-9}$. The numerical integration range for $q$ in Eqs. (\ref{P22k}) and (\ref{P13k}) is $10^{-6} \leq q \leq 10^{2}$. In this analysis, we use the normalization of $A$ defined in Eq.(\ref{Pka}) instead of $\sigma_8$. This is due to the fact that the specific inflation model predicts the specific value of $A$. However, $\sigma_8$ value is affected by the secondary effects like DE as shown in the table \ref{tab1}.

%%%%%%%%%%%%%%%%%%%%%%%%%%%%%%%%%%%%%%%%%%%%%%
\begin{figure}
\centering
\vspace{1.5cm}
  %\begin{table}%
    \setlength{\tabcolsep}{1pt} % General space between cols (6pt standard)
    \renewcommand{\arraystretch}{0.05} % General space between rows (1 standard)
\begin{tabular}{ccc}
\epsfig{file=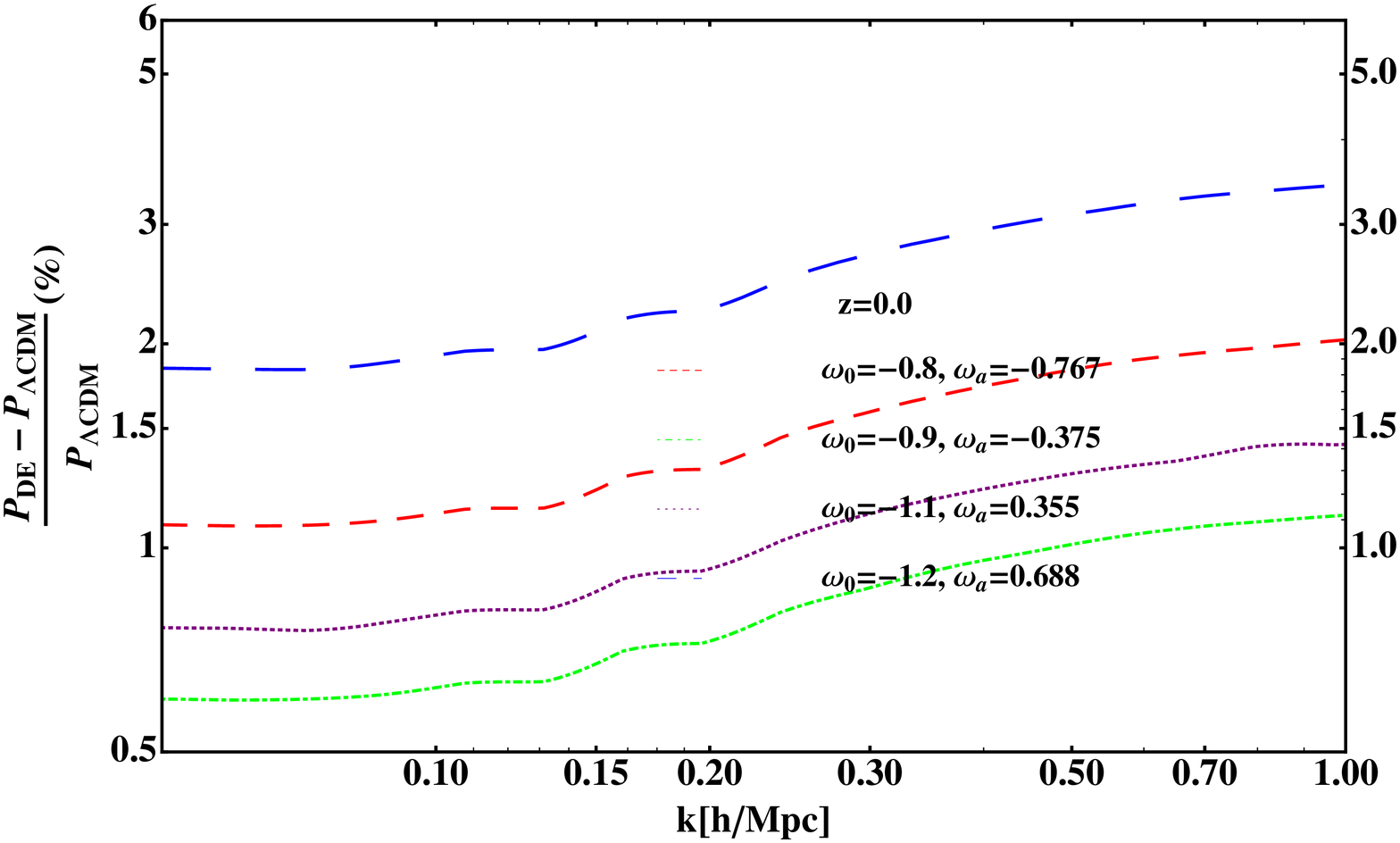,width=0.33\linewidth,clip=} &
\epsfig{file=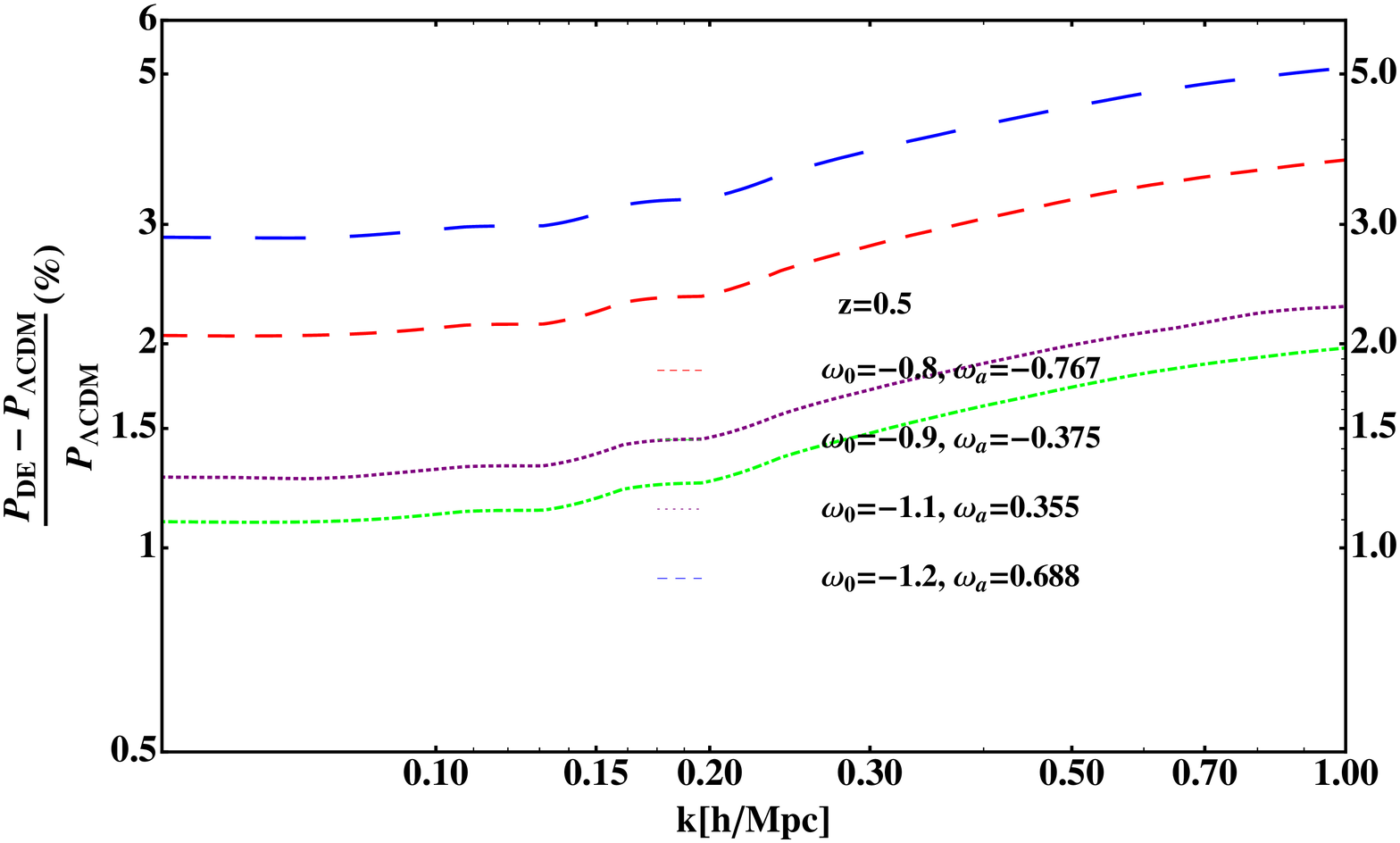,width=0.33\linewidth,clip=} &
\epsfig{file=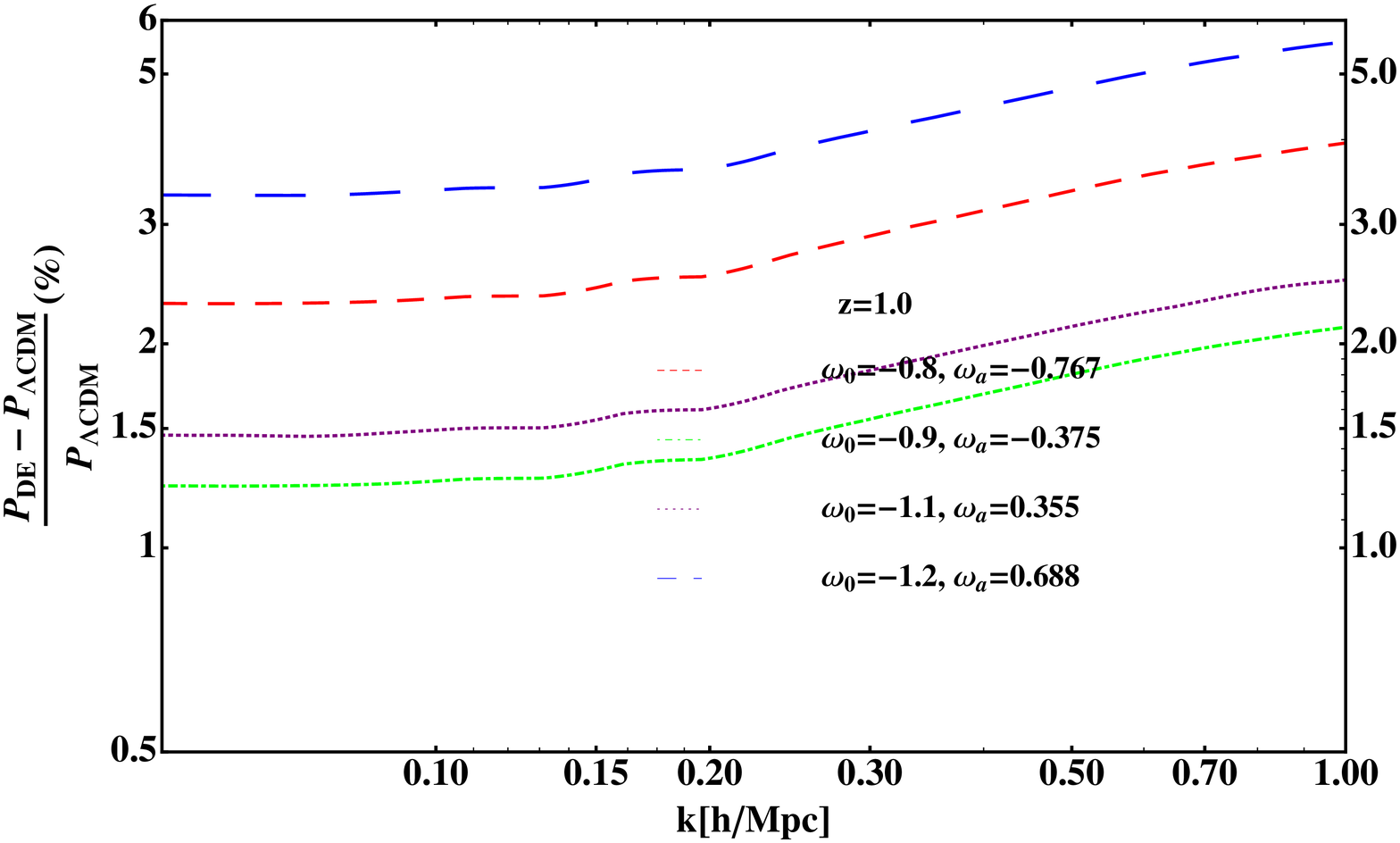,width=0.33\linewidth,clip=}\\
\end{tabular}
%\egroup
\vspace{-0.5cm}
\caption{The difference of matter power spectrum including one-loop correction between DE models and $\Lambda$CDM at different epoches. 1) Differences of matter power spectrum $P$ between DE models and the $\Lambda$CDM at $z=0$. The notation is M8 (dashed), M9 (dot-dashed), M11 (dotted), and M12 (long-dashed), respectively. 2) Differences of $P$ at $z=0.5$. 3) Difference of $P$s at $z=1.0$.
} \label{fig3}
%\end{table}
\end{figure}
%%%%%%%%%%%%%%
Dark energy dependence on the one-loop matter power spectrum is depicted in Fig.\ref{fig3}. We compare the matter power spectrum of each model with the one of $\Lambda$CDM. In the first column of Fig.\ref{fig3}, we show the differences of $P$ between models at the present epoch. The differences of $P$ between M8 (M9, M11, M12) and $\Lambda$CDM are 1.1 (0.6, 0.8, 1.9) \% for $k = 0.1 \hMpc$ mode and 1.8 (0.9, 1.2, 3.0) \% for $k = 0.4 \hMpc$ one. In the second column of the figure, the differences of $P$ between models at $z = 0.5$ are shown. The differences of $P$ between M8 (M9, M11, M12) and $\Lambda$CDM are 2.1 (1.1, 1.3, 3.0) \% for $k = 0.1 \hMpc$ and 3.0 (1.6, 1.9, 4.2) \% for $k = 0.4 \hMpc$. The last column show the differences of $P$ between models at $z = 1.0$. The differences of $P$ between M8 (M9, M11, M12) and $\Lambda$CDM are 2.4 (1.2, 1.5, 3.4) \% for $k = 0.1 \hMpc$ and 3.3 (1.7, 2.0, 4.5) \% for $k = 0.4 \hMpc$.

We summarize the RSD and matter power spectra result in table \ref{tab2}. We define
$\Delta f = \fr{f_{\DE}-f_{\Lambda \cdm}}{f_{\Lambda \cdm}} \times 100 (\%)$ and $\Delta P = \Bigl| \fr{P_{\DE}-P_{\Lambda \cdm}}{P_{\Lambda \cdm}} \Bigr| \times 100 (\%)$. Both $\Delta f$ and $\Delta P$ have the similar sensitivity on $\omega$ to separate DE models from $\Lambda$CDM at $z=0.25$. However, $\Delta P$ can be used for almost entire epochs to distinguish DE models. As $z$ increases, do does $\Delta P$. This is not able to be achieved by $RSD$. Also the bigger the departure of $\oo$ value from -1, the larger the $\Delta P$.

%%%%%%%%%%%%%%%%%%%%%%%%%%%%%%%%%%%%%%%%%%%%%%%%%%
\begin{center}
\begin{table}
\begin{tabular}{ |c||c|c|c|c|c|c|c|c|  }
\hline
 &\multicolumn{6}{c|}{Matter Power Spectrum $\Delta P$ (\%)} & \multicolumn{2}{c|}{RSD $\Delta f$ (\%)} \\ \cline{2-9}
Models & \multicolumn{2}{c|}{z=0.0} &  \multicolumn{2}{c|}{z=0.5} &\multicolumn{2}{c|}{z=1.0} & z=0 & z=0.25 \\ \cline{2-7}
 & $k =$ 0.1 & 0.4 & 0.1 & 0.4 & 0.1 & 0.4 &  & \\
\hline
M8 & 1.1 & 1.8 & 2.1 & 3.0 & 2.4 & 3.2 & 0.2 & -2.3 \\
M9 & 0.6 & 0.9 & 1.1 & 1.6 & 1.2 & 1.7 & 0.2 & -1.2 \\
M11 & 0.8 & 1.2 & 1.3 & 1.9 & 1.5 & 2.0 & 0.2 & 1.2 \\
M12 & 1.9 & 3.0 & 3.0 & 4.2 & 3.4 & 4.5 & 0.2 & 2.4 \\
\hline
\end{tabular}
\caption{Summary of results in $\Bigl| \fr{P_{\DE}-P_{\Lambda \cdm}}{P_{\Lambda \cdm}} \Bigr|$ and $\fr{f_{\DE}-f_{\Lambda \cdm}}{f_{\Lambda \cdm}}$. $k$ in unit of $\hMpc$.}
\label{tab2}
\end{table}
\end{center}
%%%%%%%%%%%%%%%%%%%%%%%%%%%%%%%%%%%%%%%%%%%%%%

%%%%%%%%%%%%%%%%%%%%%%%%%%%%%%%%%%%%%%%%%%%%%%%%%%%%%%%%%%%%%%%%%%%%%%%%%
\section{Conclusions}
\setcounter{equation}{0}
%%%%%%%%%%%%%%%%%%%%%%%%%%%%%%%%%%%%%%%%%%%%%%%%%%%%%%%%%%%%%%%%%%%%%%%%%

Cosmic microwave background is degenerated for the different dark energy models even if one fix the other cosmological parameters. This degeneracy can be broken when one combine CMB with LSS. If we parameterize the dark energy equation of state by CPL, then we can find the various combination of ($\oo, \oa$) which can produce the same angular acoustic scale for each other. These models produce the different prediction for the growth rate which can be determined by the galaxy redshift space distortions. However, the growth rate is scale independent measurement and the differences between models can be reached the maxima at the specific epoch, like $z \simeq 0.25$. Even in this case, maximum difference is about 6 \% for the considered models. If we consider the matter power spectrum including the one-loop correction, then the model dependence on the matter power spectrum is increased. If the accuracy of the future galaxy survey is reached to 5 \%, then one can rule out many dark energy models which are degenerated by CMB and RSD.

%%%%%%%%%%%%%%%%%%%%%%%%%%%%%%%%%%%%%%%%%%%%%%%%%%%%%%%%%%%%%%%%%%%%%%%%
\section*{Acknowledgments}
%%%%%%%%%%%%%%%%%%%%%%%%%%%%%%%%%%%%%%%%%%%%%%%%%%%%%%%%%%%%%%%%%%%%%%%%%
This work were carried out using computing resources of KIAS Center for Advanced Computation. S.L would like to thank for the hospitality at APCTP during the program TRP.

%%%%%%%%%%%%%%%%%%%%%%%%%%%%%%%%%%%%%%%%%%%%%%%%%%%%%%%%%%%%%%%%%%%%%%%%
%\appendix
\renewcommand{\theequation}{A-\arabic{equation}}
% redefine the command that creates the equation no.
\setcounter{equation}{0}  % reset counter
\section*{APPENDIX}  % use *-form to suppress n1umbering
%*\section*{Appendix}
%\numberwithin{equation}{section}
%\setcounter{equation}{0}
%%%%%%%%%%%%%%%%%%%%%%%%%%%%%%%%%%%%%%%%%%%%%%%%%%%%%%%%%%%%%%%%%%%%%%%%%

One can expand the CMB degenerated DE models. One is able to find the combinations of ($\oo,\oa$) to produce the same value of $\theta_{s}(z_{\ast})$ for all models. If one just considers $\theta_{s}(z_{\ast})$, then one can keep lowering the value of $\oo$ to find the proper $\oa$. We obtain corresponding $\oa$ values as given in the Table. \ref{tab3} by varying $\oo$. We label each model ($\oo,\oa$) = (-0.5,-2.035), (-0.6,-1.599), (-0.7,-1.176), (-0.8,-0.767), (-0.9,-0.375), (-1.1,0.355), (-1.2,0.688), (-1.3,0.993), (-1.4,1.266), and (-1.5,1.502) as M5, M6, M7, M8, M9, M11, M12, M13, M14, and M15, respectively.

However, CMB also provides the accurate measurements on its acoustic heights. From M5 to M14, the differences in the CMB angular power spectra between models are sub percent level for all scales. However, the differences of the CMB angular power spectrum between M15 and $\Lambda$CDM become more than 1 \% for almost all scales. Thus, one can put the lower limit on $\oo$ value ({\it i.e.} $\oo > -1.5$). This can be used as a useful prior in other observation like SNe Ia. One might be able to rule out phantom models with $\oo > 1.5$ if we use the same cosmological parameters as $\Lambda$CDM. M15 also produces about the 9 \% deviation in $\sigma_8$ value from that of $\Lambda$CDM. If the accuracy of RSD observation at $z=0.25$ reaches to 3 \% level, then one can distinguish M5, M6, M7, M13, and M14 from $\Lambda$CDM.

If one further considers the matter power spectrum, then one can have the stronger constraint on $\oo$. The difference of linear matter power spectra at the present epoch between M5 (M6, M7, M8, M9, M11, M12, M13, M14, M15) and $\Lambda$CDM becomes 2.2 (1.9, 1.5, 1.1, 0.6, 0.8, 1.9, 3.6, 7.3, 16.9) \% at $k = 0.1 \hMpc$. Thus, if the accuracy of the galaxy redshift survey reaches to 5 \% level, then both M14 and M15 can be ruled out. If one considers the matter power spectrum including the one-loop correction, then one can have the even stronger constraints on $\omega$. We consider the matter power spectrum including one-loop correcting using SPT at z = 1.0. The difference of the matter power spectra between M5 (M6, M7, M8, M9, M11, M12, M13, M14, M15) and $\Lambda$CDM becomes 4.85 (4.12, 3.28, 2.33, 1.25, 1.49, 3.37, 5.97, 10.51, 20.83) \% at $k = 0.1 \hMpc$. Also if we consider the scale $k =0.4 \hMpc$, then it becomes 6.59 (5.58, 4.44, 3.14, 1.69, 1.99, 4.48, 7.87, 13.71, 26.50) \%. Thus, even M13 will be rule by the 5 \% level accuracy measurement. These are summarized in table .\ref{tab3}.

%%%%%%%%%%%%%%%%%%%%%%%%%%%%%%%%%%%%%%%%%%%%%%%%%%
\begin{center}
\begin{table}
\begin{tabular}{ |c||c|c|c|c|c|c|c|  }
\hline
Model & ($\oo, \oa$) & $\sigma_8$ & $\Delta \sigma_8$ & \multicolumn{2}{c|}{$\Delta f$} & \multicolumn{2}{c|}{$\Delta P(z=0.5)$ (\%)} \\ \cline{5-8}
      &              &            &  & $z = 0$ & $z=0.25$ & $k=0.1$ & $k=0.4$ \\ \cline{5-8}
\hline
M5 & (-0.5,-2.035) & 0.855 & 1.183 & -0.42 & -5.56 & 4.85 & 6.59  \\
M6 & (-0.6,-1.599) & 0.853 & 0.947 & -0.35 & -4.51 & 4.12 & 5.58  \\
M7 & (-0.7,-1.176) & 0.852 & 0.828 & -0.27 & -3.43 & 3.28 & 4.44 \\
M8 & (-0.8,-0.767) & 0.850 & 0.592 & -0.19 & -2.32 & 2.33 & 3.14  \\
M9 & (-0.9,-0.375) & 0.848 & 0.355 & -0.10 & -1.18 & 1.25 & 1.69  \\
M10 & (-1.0,0)     & 0.845 & 0     & 0     & 0     & 0    & 0  \\
M11 & (-1.1,0.355) & 0.842 & -0.355 & 0.12 & 1.22 & 1.49 & 1.99 \\
M12 & (-1.2,0.688) & 0.837 & -0.947 & 0.23 & 2.50 & 3.37 & 4.48  \\
M13 & (-1.3,0.993) & 0.830 & -1.775 & 0.36 & 3.84 & 5.97 & 7.87 \\
M14 & (-1.4,1.266) & 0.814 & -3.669 & 0.52 & 5.26 & 10.51 & 13.71  \\
M15 & (-1.5,1.502) & 0.770 & -8.876 & 0.70 & 6.77 & 20.83 & 26.50  \\
\hline
\end{tabular}
\caption{Summary of results in $\Delta f$ and $|\Delta P|$ at z = 0.5. $k$ in unit of $\hMpc$.}
\label{tab3}
\end{table}
\end{center}
%%%%%%%%%%%%%%%%%%%%%%%%%%%%%%%%%%%%%%%%%%%%%%

\end{document}